\documentclass[12pt]{article}
\usepackage{a4wide}
\usepackage{amssymb}
\usepackage{graphicx}
\begin{document}

\newcommand{\md}{{\mathrm{d}}}

{\renewcommand{\thefootnote}{\fnsymbol{footnote}}
\begin{center}
{\LARGE  Harmonic cosmology: How much can we know\\
about a universe before the big bang?}\\
\vspace{1.5em}
Martin Bojowald\footnote{e-mail address: {\tt bojowald@gravity.psu.edu}}
\\
\vspace{0.5em}
Institute for Gravitation and the Cosmos,\\
The Pennsylvania State
University,\\
104 Davey Lab, University Park, PA 16802, USA\\
\vspace{1.5em}
\end{center}
}

\setcounter{footnote}{0}

\begin{abstract}
 Quantum gravity may remove classical space-time singularities and
 thus reveal what a universe at and before the big bang could be
 like. In loop quantum cosmology, an exactly solvable model is
 available which allows one to address precise dynamical coherent
 states and their evolution in such a setting. It is shown here that
 quantum fluctuations before the big bang are generically unrelated to
 those after the big bang. A reliable determination of pre-big bang
 quantum fluctuations would require exceedingly precise observations.
\end{abstract}

\section{Introduction}

Cosmology, as the physics of the universe as a whole, places special
limitations on scientific statements to be reasonably inferred from
observations. On a general basis, this was discussed in
\cite{Uncertain} who noted that some general properties seem to arise
automatically during cosmological evolution. Their verification for
our universe then does not reveal anything about its possibly more
complicated past. For instance, isotropization \cite{Isotropize}
demonstrates that the observation of a nearly isotropic present
universe does not present much useful information on an initial state
at much earlier times. Similarly, one may add decoherence to the list
which
can be seen as doing the same for the observation of a nearly
classical present universe: Most initial states would decohere and
arrive at a semiclassical state; thus its observation does not rule
out stronger quantum behavior at earlier stages. Given this, it is
surprising to see recent discussions about a universe before the big
bang in several approaches to quantum gravity. Not just statements
about the possibility of the universe having existed before the big
bang are being made, which could in principle be inferred from a
general analysis of equations of motion, but even assumptions on its
classicality or claims about the form of its state such as its
fluctuations at those times are put forward. Even if this may be
possible theoretically, it raises the question how much one can really
learn about a pre-big bang universe from observations.

A solvable model of quantum cosmology is used here to draw
cosmological conclusions within this model about the possible nature
of the big bang and a universe preceding it. The viewpoint followed
has been spelled out in \cite{BeforeBB} where also some of the results
have already been reported. This model does show a bounce at small
volume instead of the classical singularity present in solutions of
general relativity. We therefore use it to shed light on the general
questions posed above. The promotion of this specific model is not
intended as a statement that its bounce would be generic for quantum
gravity even within the same framework. In fact, the conclusion of the
bounce in this and related models available so far is based on several
specific properties which prevent a generic statement about this form
of singularity removal. We rather take the following viewpoint: Assume
there is a theoretical description of a bouncing universe; what
implications can be derived for its pre-bounce state?

The specific model used is distinguished by a key fact which makes it
suitable for addressing such a general question: As a solvable model,
the system displays properties of its dynamical coherent states which,
to some degree, are comparable to those of the harmonic
oscillator. There is no back-reaction of quantum fluctuations or
higher moments of a state on the time evolution of its expectation
values. Their trajectory is thus independent of the spreading of a
state if it occurs, a property which is well known for the harmonic
oscillator (for instance, as a simple consequence of the Ehrenfest
theorem).

This behavior is certainly special compared to other quantum systems,
but the precise derivation of properties of dynamical coherent states
it allows are nevertheless important. For instance, a large part of
theoretical quantum optics is devoted to a computation of fluctuations
in squeezed coherent states of the harmonic oscillator. Similarly, the
solvable model studied here allows detailed calculations of its
dynamical coherent states which are of interest for quantum
cosmology. In particular, the solvable model we will be using
eliminates the classical big bang singularity by quantum
effects. Dynamical coherent states highlight the behavior of
fluctuations of the state of the universe before and after the big
bang.

Care is, however, needed for the physical interpretation of the
results. While quantum optics allows one to prepare a desired state
and perform measurements on it, quantum cosmology has to make use of
the one universe state that is given to us. Unlike quantum optics,
where states can be prepared to be close to harmonic oscillator
states, we cannot realize states of the solvable quantum cosmological
model. The real universe is certainly very different from anything
solvable, and so the availability of a certain feature or numerical
result in a solvable model is unlikely to be related to a realistic
property of the universe. Thus, as spelled out in \cite{BeforeBB}, we
focus on pessimism in our analysis of the solvable model: the {\em
  in}ability of making certain predictions even in a fully controlled
model is likely to be a reliable statement much more generally; adding
complications of a real universe would only make those predictions
even more difficult.

For solvable systems of this kind it is easier by far to solve
equations of motion for expectation values and fluctuations directly,
rather than taking the detour of a specifically represented wave
function. Properties of coherent states are then determined by
selecting solutions of fluctuations which saturate uncertainty
relations. We will first illustrate this procedure for the harmonic
oscillator with an emphasis on squeezed states.
We present this brief review in section \ref{Squeezed} intended as an
introduction of the methods then used in quantum cosmology.

The main part of this paper is an application of those methods to the
solvable system of quantum cosmology, as well as a physical
discussion. Instead of different cycles of a harmonic oscillator, in
quantum cosmology we will be dealing with the pre- and post-big bang
phase of a universe. Just as fluctuations in a squeezed harmonic
oscillator state can oscillate during the cycles, fluctuations in
quantum cosmology may change from one phase to the next. Our main
concern will be the reliability of predictions about the precise state
of the universe before the big bang, based on knowledge we can achieve
after the big bang.

\section{Squeezed states}
\label{Squeezed}

From the harmonic oscillator Hamiltonian
$\hat{H}=\frac{1}{2m}\hat{p}^2+ \frac{1}{2}m\omega^2\hat{q}^2$ we
have equations of motion
\begin{eqnarray} \label{hoeom}
 \frac{{\rm d}}{{\rm d} t}\langle\hat{q}\rangle =
\frac{1}{i\hbar}\langle[\hat{q},\hat{H}]\rangle= \frac{1}{m}
\langle\hat{p}\rangle \quad,\quad
 \frac{{\rm d}}{{\rm d} t}\langle\hat{p}\rangle =
\frac{1}{i\hbar}\langle[\hat{p},\hat{H}]\rangle
 =-m\omega^2\langle\hat{q}\rangle
\end{eqnarray}
for expectation values of a state. These are already in closed form
and can thus be solved without knowing anything about fluctuations or
other moments of the state. One obtains a trajectory in exact
agreement with the classical one.

To find out more about the behavior of the state while its expectation
values follow the classical trajectory, we can derive and solve
equations of motion
\begin{eqnarray}
  \frac{{\rm d}}{{\rm d}t} (\Delta q)^2&=&
 \frac{\langle[\hat{q}^2,\hat{H}]\rangle}{i\hbar}-
2\langle\hat{q}\rangle\frac{{\rm d}\langle\hat{q}\rangle}{{\rm d}t}= 
\frac{2}{m}C_{qp}\label{deltaqdot}\\
\frac{{\rm d}}{{\rm d}t} C_{qp} &=& -m\omega^2(\Delta
q)^2+\frac{1}{m}(\Delta p)^2 \label{Cqpdot}\\
\frac{{\rm d}}{{\rm d}t} (\Delta p)^2&=& -m\omega^2C_{qp} \label{deltapdot}
\end{eqnarray}
for fluctuations. Also these equations form a closed set, and can thus
be solved without requiring knowledge of higher moments which we write
in the general form
\begin{equation}
 G^{O_1\cdots O_n} := \left\langle \left((\hat{O}_1-\langle\hat{O}_1\rangle)
 \cdots (\hat{O}_n-\langle\hat{O}_n\rangle)\right)_{\rm Weyl}\right\rangle
\end{equation}
as expectation values of Weyl ordered products of operators.  It is
clear that to second order fluctuations $G^{qq}=(\Delta q)^2$ and
$G^{pp}=(\Delta p)^2$ as well as the covariance $G^{qp}=C_{qp}:=
\frac{1}{2}\langle \hat{q}\hat{p}+\hat{p}\hat{q}\rangle
-\langle\hat{q}\rangle\langle\hat{p}\rangle$ result. We thus determine
exact state properties in this way, and there is no need for an
approximation. Approximations would, however, be necessary for
anharmonic systems where equations of all the moments and of
expectation values are coupled with each other.  Systematic treatments
\cite{EffAc,Karpacz} give rise to effective equations for the
motion of expectation values including quantum corrections from
back-reaction of moments on the expectation values.

To select coherent states, we need to saturate the uncertainty
relation
\begin{equation} \label{standarduncert}
 G^{qq}G^{pp}- (G^{qp})^2\geq \frac{\hbar^2}{4}\,.
\end{equation}
Among all coherent states, those with vanishing correlations,
$G^{qp}=0$, have minimal uncertainties. For harmonic oscillator
states, this requires $\Delta p=m\omega\Delta q$ for $G^{qp}$ to
remain zero at all times according to (\ref{Cqpdot}). As the equations
of motion (\ref{deltaqdot}) and (\ref{deltapdot}) then show, the
uncertainties must be constant in time: The state does not spread at
all while its expectation values follow the classical trajectory. If
we require the uncertainty relation to be saturated, fluctuations are
determined uniquely for uncorrelated states. The corresponding state
is the harmonic oscillator ground state, where $\Delta q=
\sqrt{\hbar/2m\omega}$ and $\Delta p=\sqrt{m\omega \hbar/2}$.  But for
non-zero correlations there are differently squeezed states whose
relation between $\Delta q$ and $\Delta p$ does not agree with that
realized for the ground state, even if saturation of the uncertainty
relation is required.

More generally squeezed states exist if we remain on the saturation
surface but allow non-zero covariance. With $G^{qp}\not=0$,
fluctuations now depend on time in an oscillatory manner. These
properties can all be derived without using a specific representation
of states. For comparison, one may easily check that states saturating
the uncertainty relation are Gaussians $\psi(q)=
\exp(-z_1q^2+z_2q+z_3)$ in the position representation, with ${\rm Re}
z_1>0$ for normalizability. If $z_1$ is real, we have uncorrelated
states at the absolute minimum of the saturation surface. Otherwise,
the covariance is given by $G^{qp}=-\frac{{\rm Im} z_1}{2{\rm Re}
  z_1}\hbar$.

\section{Harmonic quantum cosmology}

Isotropic cosmology is based on one pair of canonical variables: the
scale factor $a$ which determines the spatial size of a universe and
its momentum, related to $\md a/\md\tau$ with proper time
$\tau$. Dynamics is determined by the Friedmann equation
\begin{equation} \label{Friedmann}
\left( \frac{1}{a}\frac{\md a}{\md\tau}\right)^2 = \frac{8\pi G}{3}\rho
\end{equation}
where the energy density $\rho$ of matter provides the source and $G$
is Newton's constant. We will be especially interested in a matter
source given by a free, massless scalar $\phi$ with momentum
$p_{\phi}$. Its energy density is independent of $\phi$ and takes the
purely kinetic form 
\begin{equation} \label{rho}
 \rho=\frac{p_{\phi}^2}{2a^6}\,.
\end{equation}
As we will see, this system provides a solvable model playing the role
of the harmonic oscillator for quantum cosmology.

Since our aim is to study properties of dynamical coherent states
especially in their relation between pre- and post-big bang phases, we
use a quantization which eliminates the classical singularity
occurring when solutions reach $a=0$. In isotropic models, this can
generally be achieved in loop quantum cosmology \cite{LivRev},
where geometry is quantized and the spectrum of geometrical operators
becomes discrete. The form of the discreteness, i.e.\ whether it is
the scale factor itself whose spectrum is equidistantly spaced or
rather some of its powers such as the volume $a^3$, is to be
determined from the precise form of dynamics in quantum gravity.

Fortunately, our model will not be sensitive to this detail,
and we can thus leave the precise form of discreteness open. We
introduce new canonical variables
\begin{equation}
 v=\frac{1}{1-x} a^{2(1-x)} \quad,\quad P=a^{2x}\dot{a}
\end{equation}
where the quantization of $v$ will have an equidistant spectrum. As
$x$ is changed between different models, the form of the spectrum for
geometrical quantities takes on all possibilities. From loop quantum
gravity one expects $x$ to lie within the range $0<x<-1/2$
\cite{InhomLattice}, with phenomenological arguments favoring a value
near $-1/2$
\cite{APSII,SchwarzN,RefinementMatter,RefinementInflation,Vector,Tensor}.

A further implication of the discreteness of $\hat{v}$ in a
quantization is that no operator for $P$ itself can exist. This is
similar to quantization on a circle, where the angle as a local
coordinate cannot be promoted to an operator but only its periodic
functions which are global functions on the configuration
space. Similarly, there is no operator for $P$ or $\dot{a}$ in loop
quantum cosmology, but only quantizations of $\sin \delta P$ and
$\cos\delta P$ for real $\delta$. Notice that the periodicity is kept
free since operators for all $\delta$ exist. This happens because the
space of values $\dot{a}$ is not simply modified to a circle by
postulating a periodic identification. The quantum configuration
space \cite{Bohr} is rather a compactification of the
real line, the so-called Bohr compactification, which contains the
whole real line as a dense set but is itself compact. Accordingly, an
orthonormal basis can be written as $\{|\mu\rangle:\mu\in{\mathbb
R}\}$ for this non-separable Hilbert space. The basic operators act as
\begin{equation}
 \hat{v}|\mu\rangle = \hbar\mu|\mu\rangle \quad,\quad
 \widehat{\exp(i\delta P)}|\mu\rangle = |\mu+\delta\rangle
\end{equation}
with commutator
\[
 [\hat{v},\widehat{\exp(i\delta P)}] = \hbar\widehat{\exp(i\delta P)}
\]
as appropriate for a canonical pair $(v,P)$.

Thus, also the quantized Friedmann equation must make use of
$\sin\delta P$ rather than $\dot{a}$ directly. At this point, $\delta$
would be fixed such that dynamics can be restricted to a separable
sector of the full Hilbert space. Qualitative properties discussed
here do not depend on the value of $\delta$, nor on that of $x$ used
in the definition of classical variables. The construction of loop
quantum cosmology shows that the term $\dot{a}^2a^4$, which as per
equation (\ref{Friedmann}) is proportional to $p_{\phi}^2$ for a free
scalar, then takes the form $v^2\sin^2P$. Independently of $x$, it
reduces to the classical expression in the small-curvature limit $P
\ll 1$. (We do not write this as a proper operator yet because we will
later be led to a specific factor ordering which is being kept free
for now.)  Disregarding numerical factors, the dynamical equation
combining (\ref{Friedmann}) and (\ref{rho}) can thus be written in the
form
\begin{equation} \label{pphi}
 p_{\phi} = |v\sin P|\,.
\end{equation}
If the matter value $\phi$ is interpreted as a time variable, this
shows that the Hamiltonian in this internal time description is
$H=|v\sin P|$. For small $P$ and disregarding the absolute value, this
would be a quadratic Hamiltonian and we would have the same decoupling
behavior as in the harmonic oscillator. (This Berry--Keating--Connes
Hamiltonian \cite{BerryKeating,Connes} plays a role in
the context of the Riemann hypothesis.)  However, the sine certainly
makes the expression non-quadratic in canonical variables.

Remarkably, the system is still solvable even with the sine. We have
to change variables to non-canonical ones and specify a precise factor
ordering of the Hamiltonian. Let us introduce the operator $\hat{J}:=
\hat{v}\widehat{\exp(iP)}$ which has a well-defined representation in
loop quantum cosmology thanks to its periodicity in $P$. Together
with $\hat{v}$, we have the linear algebra
\begin{equation}\label{comm}
  [\hat{v},\hat{J}]=\hbar\hat{J}\quad,\quad {}
  [\hat{v},\hat{J}^{\dagger}]=-\hbar\hat{J}^{\dagger}\quad,\quad {}
  [\hat{J},\hat{J}^{\dagger}]=-2\hbar\hat{v}-\hbar^2
\end{equation}
equivalent to an ${\rm sl}(2,{\mathbb R})$ algebra with a trivial
central extension by $-\hbar^2$. These operators quantize a complete
set of phase space functions and can thus be used as basic operators
for the construction of a quantum theory. They are also suitably
chosen for the dynamics since the combination
$\hat{H}=\frac{1}{2i}(\hat{J}-\hat{J}^{\dagger})$ is a quantization of
the classical Hamiltonian (\ref{pphi}) in a specific factor
ordering. (Note the absence of the absolute value compared to
(\ref{pphi}). As shown in \cite{BounceCohStates}, this does not matter
for states which satisfy $\Delta H\ll \langle\hat{H}\rangle$ as it
would be part of the conditions for semiclassical states.)  Most
importantly, the Hamiltonian is a linear combination of two of our
basic operators, which makes the system solvable and decouples
fluctuations from expectation values: in equations of motion as in
(\ref{hoeom}), $[\hat{O},\hat{H}]$ is linear in basic operators for
any basic operator $\hat{O}\in\{\hat{v},\hat{J},\hat{J}^{\dagger}\}$.

With complex basic variables we have to impose a reality condition
making sure that $\hat{J}\hat{J}^{\dagger}=\hat{v}^2$. Taking an
expectation value of this relation,
\begin{equation} \label{reality}
 |\langle\hat{J}\rangle|^2-(\langle\hat{v}\rangle+
{\textstyle\frac{1}{2}}\hbar)^2=G^{vv}-G^{J\bar{J}}+
 \frac{1}{4}\hbar^2\,,
\end{equation}
shows that fluctuations and expectation values are related in a
specific way. But this does not involve a coupling of all infinitely
many moments. Moreover, one can impose (\ref{reality}) after having
solved the equations of motion. Thus, no severe complication to
solvable systems in real variables arises.

We can thus efficiently make use of the same solution procedure
introduced above for the harmonic oscillator, and determine properties
of dynamical coherent states. Equations of motion for expectation
values, fluctuations and covariances are
\begin{equation} \label{vJdot}
 \frac{\md}{\md t}\langle\hat{v}\rangle=
 \frac{1}{i\hbar}\langle[\hat{v},\hat{H}]\rangle=
 -\frac{1}{2}(\langle\hat{J}\rangle+\langle\hat{J}^{\dagger}\rangle)
 \quad,\quad \frac{\md}{\md t}\langle\hat{J}\rangle=
 \frac{1}{i\hbar}\langle[\hat{J},\hat{H}]\rangle=
 -\langle\hat{v}\rangle-\frac{1}{2}\hbar=\frac{\md}{\md
 t}\langle\hat{J}^{\dagger}\rangle\,.
\end{equation}
and
\begin{eqnarray}
 \dot{G}^{vv} &=& -G^{vJ}-G^{v\bar{J}} \label{Gvvdot}\\ 
\dot{G}^{JJ} &=& -2G^{vJ}\quad,\quad
 \dot{G}^{\bar{J}\bar{J}} = -2G^{v\bar{J}}\\ \dot{G}^{vJ} &=&
 -\frac{1}{2} G^{JJ}-\frac{1}{2}G^{J\bar{J}}-G^{vv} \quad,\quad
 \dot{G}^{p\bar{J}} = -\frac{1}{2} G^{\bar{J}\bar{J}}-
 \frac{1}{2}G^{J\bar{J}}-G^{vv}\\ \dot{G}^{J\bar{J}} &=&
 -G^{vJ}-G^{v\bar{J}} \label{GJbarJdot}\,.
\end{eqnarray}

All equations can easily be solved: For expectation values we obtain,
taking into account the reality condition (\ref{reality}),
\begin{eqnarray}
  \langle\hat{v}\rangle(\phi) &=& \frac{1}{2}(Ae^{-\phi}
  +Be^{\phi})-\frac{1}{2}\hbar\\
 \langle\hat{J}\rangle(\phi) &=& \frac{1}{2}(Ae^{-\phi}
 -Be^{\phi})+i H
\end{eqnarray}
where $H:=\langle\hat{H}\rangle$. For these solutions, the reality
condition implies $AB=H^2+c_1-\frac{1}{4}\hbar^2$ with
$c_1:=G^{J\bar{J}}-G^{pp}$ which according to (\ref{Gvvdot}) and
(\ref{GJbarJdot}) is constant in time. For states which are
semiclassical at large volume (or just at one time) we must have
$c_1\ll H^2$ and thus positive $AB$.  This clearly shows the absence
of the classical singularity since
$\langle\hat{v}\rangle(\phi)=H\cosh(\phi-\delta)-\frac{1}{2}\hbar$
with $e^{2\delta}=A/B$ never reaches zero. From now on we set $A=B$
without loss of generality since it simply amounts to a shift in the
origin of our internal time $\phi$.

For fluctuations and covariances we have explicit solutions
\begin{eqnarray}
 G^{vv}(\phi) &=& \frac{1}{2}(c_3e^{-2\phi}+c_4e^{2\phi})- 
\frac{1}{4}(c_1+c_2)\\
 G^{JJ}(\phi) &=& \frac{1}{2}(c_3e^{-2\phi}+c_4e^{2\phi})+
\frac{1}{4}(3c_2-c_1)-
 i(c_5e^{\phi}-c_6e^{-\phi})\\
 G^{\bar{J}\bar{J}} &=& \frac{1}{2}(c_3e^{-2\phi}+c_4e^{2\phi})+
\frac{1}{4}(3c_2-c_1)+
 i(c_5e^{\phi}-c_6e^{-\phi})\\
 G^{vJ}(\phi) &=& \frac{1}{2}(c_3e^{-2\phi}-c_4e^{2\phi})+ 
\frac{i}{2}(c_5e^{\phi}+c_6e^{-\phi})\\
 G^{v\bar{J}}(\phi) &=& \frac{1}{2}(c_3e^{-2\phi}-c_4e^{2\phi})- 
\frac{i}{2}(c_5e^{\phi}+c_6e^{-\phi})\\
 G^{J\bar{J}}(\phi) &=& \frac{1}{2}(c_3e^{-2\phi}+c_4e^{2\phi})+
 \frac{1}{4}(3c_1-c_2)
\end{eqnarray}
with constants of integration $c_1$ (see \cite{BounceCohStates} for
further details).

\section{Dynamical coherent states}

Depending on the integration constants $c_i$, there are obviously
different possibilities in the relation of fluctuations before and
after the big bang, defined as the minimum of
$\langle\hat{v}\rangle(\phi)$. For special choices one can arrange the
fluctuations to be identical at positive and negative $\phi$, but not
generically. The quantum state and its coherence will thus appear
differently before and after the big bang, depending on what property
of the state an observation might be sensitive to.

This behavior of fluctuations is a direct consequence of the
replacement of the classical singularity by a bounce, as it occurred
due to the discreteness of geometry. Without taking into account the
discreteness, we would have arrived at the Berry--Keating--Connes
Hamiltonian quantizing $vP$ directly. This system differs from the
harmonic oscillator (although it can be mapped by a canonical
transformation to an upside-down harmonic oscillator) in the fact that
it does not have constant or even bounded fluctuations in its coherent
states. However, the ratio $G^{vv}/\langle\hat{v}\rangle^2$ is exactly
preserved and ensures that an initial semiclassical state remains
semiclassical. This behavior is also true for the loop quantized model
for small curvature $P$. But near the bounce there are strong
deviations which result in a non-constant ratio
$G^{vv}/\langle\hat{v}\rangle^2$. One can see this directly from the
equations of motion (\ref{vJdot}) and (\ref{Gvvdot}), from which we
derive
\begin{equation}
 \frac{\md}{\md t}\frac{G^{vv}}{\langle\hat{v}\rangle^2} =
 -\frac{G^{vJ}+G^{v\bar{J}}}{\langle\hat{v}\rangle^2}+
 \frac{\langle\hat{J}\rangle+
\langle\hat{J}^{\dagger}\rangle}{\langle\hat{v}\rangle^3}G^{vv}\,.
\end{equation}
This is zero for $G^{vJ}+G^{v\bar{J}}= \langle\hat{v}\rangle^{-1}
(\langle\hat{J}\rangle+\langle\hat{J}^{\dagger}\rangle) G^{vv}$ or,
using the definition of $\hat{J}$, $G^{vv\cos P}=G^{vv}G^{v\cos
P}/\langle\hat{v}\rangle$. For small $P$ we can ignore the cosine and
indeed derive that $G^{vv}/v^2$ is constant. But near the bounce
deviations of $\cos P$ from its low-curvature value are important and,
in general, imply non-constant relative fluctuations through the
bounce. Another possibility to satisfy the required relation is to
have uncorrelated states such that $G^{vv\cos P}\approx G^{vv}\cos P$
and $G^{v\cos P}\approx \langle\hat{v}\rangle \cos P$ which would also
imply a constant $G^{vv}/\langle\hat{v}\rangle^2$ even through the
bounce at high curvature.  For squeezed states with correlations,
however, relative fluctuations before and after the bounce can
differ from each other, even though they are nearly constant in any
pre- or post-bounce phase.

To make this precise, we can determine explicit forms of the
uncertainty relations in our variables. For any pair
$(\hat{O}_1,\hat{O}_2)$ of self-adjoint operators we have
\begin{equation}
 G^{O_1O_1}G^{O_2O_2}-(G^{O_1O_2})^2\geq \frac{1}{4}\langle
 -i[\hat{O}_1,\hat{O}_2]\rangle^2
\end{equation}
so that in our case we can write three different inequalities:
\begin{equation}
 G^{vv}G^{J+\bar{J},J+\bar{J}}- (G^{v,J+\bar{J}})^2
 \geq \hbar^2H^2
\end{equation}
for the pair $(\hat{v},\hat{J}+\hat{J}^{\dagger})$,
\begin{equation}
 G^{vv}G^{i(J-\bar{J}),i(J-\bar{J})}- (G^{v,i(J-\bar{J})})^2
  \geq \frac{1}{4}\hbar^2(\langle\hat{J}\rangle+ 
\langle\hat{J}^{\dagger}\rangle)^2
\end{equation}
for the pair $(\hat{v},i(\hat{J}-\hat{J}^{\dagger}))$ and
\begin{equation}
 G^{J+\bar{J},J+\bar{J}}G^{i(J-\bar{J}),i(J-\bar{J})}-
 (G^{J+\bar{J},i(J-\bar{J})})^2
 \geq \hbar^2(2\langle\hat{v}\rangle+\hbar)^2
\end{equation}
for the pair $(\hat{J}+\hat{J}^{\dagger},i(\hat{J}-\hat{J}^{\dagger}))$.
Inserting solutions, this provides the relations
\begin{eqnarray}
&& 4c_3c_4-\frac{1}{4}(c_1+c_2)^2\geq \hbar^2H^2 \label{uncertI}\\
&&(c_1-c_2) (c_3e^{-2\phi}+c_4e^{2\phi})+ \frac{1}{2}(c_2^2-c_1^2)
- c_5^2e^{2\phi}- 2c_5c_6- c_6^2e^{-2\phi}\nonumber\\
&&\qquad \geq \frac{1}{4}A^2\hbar^2(e^{-\phi}-
e^{\phi})^2\label{uncertII}\\
&& 4(c_1-c_2) (c_3e^{-2\phi}+c_4e^{2\phi})-
 2(c_2^2-c_1^2)
- 4c_5^2e^{2\phi}+8c_5c_6- 4c_6^2e^{-2\phi}\nonumber\\
&&\qquad \geq  A^2\hbar^2(e^{-\phi}+e^{\phi})^2 \label{uncertIII}
\end{eqnarray}
between the integration constants.

The last two relations suppress the pre-bounce parameters of
fluctuations by exponentials of $\phi$ and thus by the total
volume. This is huge at late times and fluctuation parameters from
before the big bang can thus be ignored completely after the big
bang. In other words, these two relations do not provide any constraints
on the state before the big bang.

The first relation is different since any $\phi$-dependence dropped
out. It thus presents an uncertainty relation between the pre- and
post-big bang fluctuation parameters $c_3$ and $c_4$. The covariance
term $-\frac{1}{4}(c_1+c_2)^2$ in this relation can be seen to
represent matter fluctuations: From the reality condition together
with
\begin{equation} \label{c1minusc2}
 (\Delta H)^2 = -\frac{1}{4}(G^{JJ}+G^{\bar{J}\bar{J}})
 +\frac{1}{2}G^{J\bar{J}} = \frac{1}{2}(c_1-c_2)
\end{equation}
we have
\begin{equation} \label{c1plusc2}
 c_1+c_2 = 2A^2-2H^2-2(\Delta H)^2+ \frac{1}{2}\hbar^2\,.
\end{equation}

In (\ref{uncertI}), however, $\phi$-dependent terms dropped out due to
cancellations between the fluctuation and covariance terms in
$G^{J+\bar{J},J+\bar{J}} = 4G^{vv}+2(c_1+c_2)$ and $G^{v,J+\bar{J}}=
G^{vJ}+G^{v\bar{J}}$. If the model is changed slightly or one tries to
use real observations which must be imprecise, those terms re-enter
the game and lead to a dominant expression which involves only
post-big bang fluctuations. But even with those cancellations there is
no strong relation between pre- and post-big bang quantities, as shown
now.

That squeezing is responsible for the possible asymmetry of
fluctuations around the bounce can be seen by directly computing the
quantum variables in our basic set of operators for a Gaussian
state. If this Gaussian is not completely squeezed, corresponding to a
real $z_1$, then fluctuations before and after the bounce must be
identical. A quick way to see this is to note that there are three
independent parameters in an unsqueezed Gaussian, ${\rm Re} z_2$ and
${\rm Im} z_2$ which determine the peak position in phase space and
${\rm Re} z_1$, and two constants of motion for expectation values and
fluctuations, $H=\langle\hat{H}\rangle$ and $G^{HH}$. Comparing two
Gaussian states, one before and one after the big bang along a
dynamical trajectory, at equal volume fixes the free parameter. Thus,
the spreads must also be equal at a given volume. Switching on general
squeezing, however, introduces additional parameters and the
conclusions are weaker.

Precise conditions can be found by solving the saturation equations
explicitly. We use (\ref{uncertI}) directly, and obtain three more
equations from the $e^{\pm2\phi}$ and constant coefficients of
(\ref{uncertII}) and (\ref{uncertIII}):
\begin{eqnarray}
4c_3c_4 &=& H^2\hbar^2+\frac{1}{4}(c_1+c_2)^2 \label{SatI}\\
 (c_1-c_2)c_3 -c_6^2 &=& \frac{1}{4}A^2\hbar^2\label{SatII}\\
 (c_1-c_2)c_4 -c_5^2 &=& \frac{1}{4}A^2\hbar^2\label{SatIII}\\
   4c_5c_6 &=& A^2\hbar^2+c_2^2-c_1^2\,. \label{SatIV}
\end{eqnarray}
With (\ref{c1minusc2}) and (\ref{c1plusc2}) we can eliminate $c_1$ and
$c_2$ in favor of $A$ and $\Delta H$ together with $H$. As a first
result, subtraction of (\ref{SatII}) and (\ref{SatIII}) shows that
$v$-fluctuations can only be symmetric around the bounce, i.e.\
$c_3=c_4$, if $|c_5|=|c_6|$  \cite{BounceCohStates}. We
have to look closer to see what range of deviations between $c_3$ and
$c_4$ is allowed for coherent states.

We can directly solve (\ref{SatII}) and (\ref{SatIII}) for $c_3$ and
$c_4$ in terms of $c_5$ and $c_6$, and then insert these variables
in (\ref{SatI}). In combination with (\ref{SatIV}) to eliminate the
resulting term $c_5^2c_6^2$, we then have
\begin{eqnarray}
 c_5^2+c_6^2 &=& 4\frac{H^2}{A^2} (\Delta H)^4-
 \frac{1}{2}A^2\hbar^2+(\Delta H)^2(c_1+c_2)\\
&=& 4\left(\frac{H^2}{A^2}-\frac{1}{2}\right) (\Delta H)^4
 +2\left(A^2-H^2+\frac{1}{4}\hbar^2\right) 
(\Delta H)^2-\frac{1}{2}A^2\hbar^2\,.\nonumber
\end{eqnarray}
Adding and subtracting $2c_5c_6$ from (\ref{SatIV}), we obtain
\begin{eqnarray}
 (c_5-c_6)^2 &=& 4\left(\frac{H^2}{A^2}-1\right) (\Delta H)^4
 +4\left(A^2-H^2+\frac{1}{4}\hbar^2\right) 
(\Delta H)^2-A^2\hbar^2 =: 4\delta^2\nonumber\\
 (c_5+c_6)^2 &=& 4\frac{H^2}{A^2}(\Delta H)^4 \label{delta}
\end{eqnarray}
which finally provides
\begin{eqnarray}
 |c_3-c_4| &=& \frac{1}{2(\Delta H)^2} \sqrt{(c_5+c_6)^2
  (c_5-c_6)^2} =2\frac{H}{A}\delta\\
 &=& 2\frac{H}{A} \sqrt{\left(\frac{H^2}{A^2}-1\right) (\Delta H)^4
 +\left(A^2-H^2+\frac{1}{4}\hbar^2\right) 
(\Delta H)^2-\frac{1}{4}A^2\hbar^2}  \nonumber
\end{eqnarray}
as an explicit relation for the difference in pre- and post-big bang
fluctuations of $v$. 

This expression can easily be estimated for its order of magnitude,
noting that the uncertainty relations, especially (\ref{uncertII}),
indicate that $(\Delta H)^2\sim A\hbar$ and, together with the reality
condition, $A^2\sim H^2+O(A\hbar)$. Thus, the first term in the square
root is smallest while the remaining terms are of the order
$A^2\hbar^2$. The square root $\delta$ and thus $|c_3-c_4|$ is thus of
the same order $A\hbar$ as the fluctuation $(\Delta H)^2$. The
asymmetry vanishes for
\begin{equation} \label{DeltaHsymm}
 (\Delta H)^2 = \frac{A^2}{A^2-H^2}\frac{\hbar^2}{4} \sim A\hbar
\end{equation}
which confirms results of \cite{BounceCohStates} and is consistent
with the orders of magnitude given above.

One can write the asymmetry more directly in fluctuation variables,
providing a relation
\begin{eqnarray}
&& \left| \lim_{\phi\to-\infty} \frac{G^{vv}}{\langle\hat{v}\rangle^2}- 
\lim_{\phi\to\infty} \frac{G^{vv}}{\langle\hat{v}\rangle^2}\right| =
2\frac{|c_3-c_4|}{A^2} \label{asymm}\\
&&\qquad= 4\frac{H}{A}
\sqrt{\left(1-\frac{H^2}{A^2}+\frac{1}{4}\frac{\hbar^2}{A^2}\right)
  \frac{(\Delta H)^2}{A^2} -\frac{1}{4}\frac{\hbar^2}{A^2}+
  \left(\frac{H^2}{A^2}-1\right) \frac{(\Delta H)^4}{A^4}} \nonumber
\end{eqnarray}
which resembles estimates provided independently by \cite{ACS}. For
any value for $A$, $H$ and $\Delta H$ this number is of the small
order $\hbar/A$, and thus one could conclude that the asymmetry must
be small and all coherent states are very nearly symmetric. However,
this is not true: The order $\hbar/A$ is what one already expects for
a single relative fluctuation $\lim_{\phi\to-\infty}
G^{vv}/\langle\hat{v}\rangle^2=c_3/A^2$ since $c_3\sim A\hbar$, not
just for the difference in the asymmetry. Fluctuations themselves are
certainly small in a saturated state, and (\ref{asymm}) only gives
their magnitude but does not provide information on the asymmetry of
those small quantities.

This can easily be corrected because we have full control on our
dynamical coherent state parameters. Solving for $c_4$ from
(\ref{SatIII}) and determining $c_5^2$ from (\ref{delta}), we derive
\begin{eqnarray}
 \left|1-\frac{c_3}{c_4}\right| &=& \frac{|c_4-c_3|}{c_4}\nonumber \\
 &=& \frac{2\frac{H}{A}\delta}{\frac{\delta^2}{2(\Delta H)^2}
 \pm\frac{H}{A}\delta+ \frac{1}{2}\frac{H^2}{A^2}(\Delta H)^2+
 \frac{1}{8}\frac{A^2\hbar^2}{(\Delta H)^2}} \label{relasymm}
\end{eqnarray}
which is a measure for the asymmetry independent of the total size of
fluctuations. The sign in the denominator depends on whether $c_3<c_4$
(upper sign) or $c_3>c_4$ (lower sign).  As one can see, this vanishes
for $\delta=0$, i.e.\ when (\ref{DeltaHsymm}) is fulfilled. This is,
however, only one special case and in general with $\delta\sim
A\hbar\sim (\Delta H)^2$ one can only infer that $|1-c_3/c_4|$ is of
the order $O(1)$, not necessarily close to zero. Generically, the
asymmetry for coherent states is not restricted at all. Symmetry of
fluctuations before and after the bounce is not generic even for states
saturating the uncertainty relations.

One simple example which illustrates how large the asymmetry can be is
obtained for $\delta=\frac{H}{A}(\Delta H)^2$, in which case the first
three terms in the denominator of (\ref{relasymm}) cancel each other
for the lower sign solution. For the constants, this implies that
\[
 A^2 = H^2\frac{(\Delta H)^2}{(\Delta H)^2-\frac{1}{4}\hbar^2}+
 (\Delta H)^2\sim H^2+(\Delta H)^2
\]
and thus $c_1\sim (\Delta H)^2+\frac{1}{4}\hbar^2$ and $c_2\sim -(\Delta
H)^2+\frac{1}{4}\hbar^2$ as well as $c_3\sim
\frac{1}{8}H\hbar+2(\Delta H)^2$, $c_4\sim \frac{1}{8}H\hbar$, $c_5=0$,
$|c_6|\sim 2(\Delta H)^2$. The only property which makes this
case special is that $c_5=0$, but the other magnitudes for
fluctuations are fully acceptable. With this $\delta$, the relative
asymmetry becomes $|1-c_3/c_4|\sim 16$, which is certainly a large value.

In fact, although the denominator in (\ref{relasymm}) never becomes
zero for real $\delta$, the expression is unbounded from above for the
lower sign (and reaches arbitrarily closely to $|1-c_3/c_4|=1$ for the
upper sign). Extremely large values require large fluctuations and
thus, on the saturation surface, very high squeezing. But large values
for $c_3/c_4$ can easily be reaches under appropriate conditions on
semiclassical states. At constant $A$, the maximum of $|1-c_3/c_4|$
along varying $(\Delta H)^2$ is realized for
\[
 (\Delta H)^2 = \frac{A^2}{H^2} \left(1-\frac{H^2}{A^2}\right)+
 \frac{\hbar^2}{2(1-H^2/A^2)}
\]
which results in a value
\[
 \left|1-\frac{c_3}{c_4}\right| =
 \frac{2}{\sqrt{1+\frac{1}{4}\frac{H^2\hbar^2}{(A^2-H^2)^2}}\pm 1}\,.
\]
This is indeed unbounded for the lower sign, but would require values
of $A$ significantly larger than $H$. Since previous relations told us
that $A^2-H^2\sim \hbar$ for semiclassical behavior, $A$ should not
differ too much from $H$ and we can approximate the semiclassical
maximum of the asymmetry by the value $4/(\sqrt{5}\pm 2)$. For the
lower sign, this gives a value $|1-c_3/c_4|\approx 16.9$ (or
$|1-c_3/c_4|\approx 0.94$ for the upper sign) near the example found
above.

\begin{figure}
\begin{center}
\includegraphics[width=11cm]{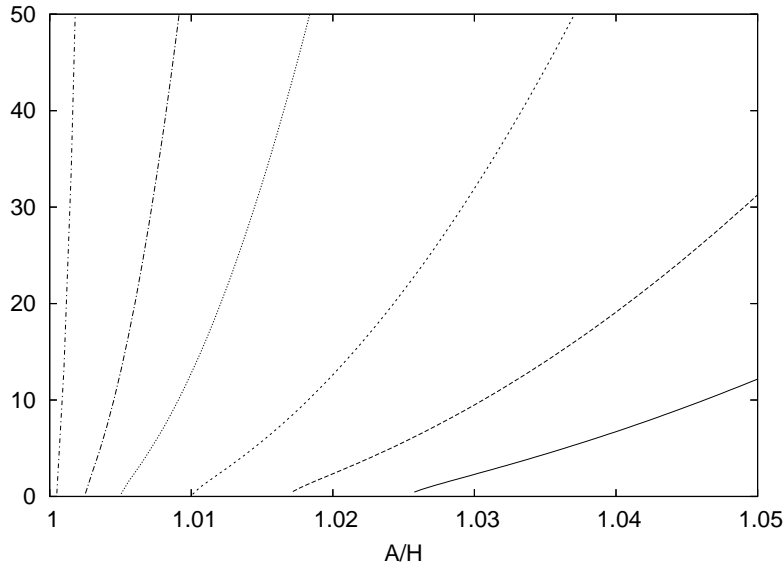}
\end{center}
\caption{The asymmetry (\ref{AsymmMin}) for minimal uncertainty
  product $c_3c_4$ as a function
  $|1-c_3/c_4|=2/((1-\frac{1}{4}\frac{\hbar^2}{H^2}x^2(x^2-
  1)^{-2})^{-1/2}-1)$ of $x=A/H$. Curves are shown for $H=10$, $H=15$,
  $H=25$, $H=50$, $H=100$ and $H=500$ with larger $H$ corresponding to
  steeper curves. \label{AsymmFluct}}
\end{figure}

One might think that a state with symmetric fluctuations makes the
product $c_3c_4$ minimal, just as a non-squeezed state does for the
usual uncertainty product in quantum mechanics according to
(\ref{standarduncert}). Interestingly, this is not necessarily the
case: (\ref{SatI}) shows that this product, for a given $H$, is
minimal for $c_1+c_2=0$, and thus $(\Delta H)^2=
A^2-H^2+\frac{1}{4}\hbar^2$. This is not the value (\ref{DeltaHsymm})
obtained above for symmetric fluctuations, unless we specialze the
parameters further to $A^2-H^2\sim\frac{1}{2}H\hbar$. More generally,
we rather have
\[
 \delta^2= \frac{H^2}{A^2}(\Delta H)^4 -\frac{1}{4}A^2\hbar^2
\]
and thus
\begin{equation} \label{AsymmMin}
  \left|1-\frac{c_3}{c_4}\right|= 
\frac{2\frac{H}{A}\delta}{\frac{H^2}{A^2}(\Delta H)^2 \pm \frac{H}{A}\delta}
\sim \frac{2}{(1-\frac{1}{4}A^2\hbar^2 (A^2-H^2)^{-2})^{-1/2} \pm 1}
\end{equation}
which can take values of a magnitude similar to those obtained above.
It vanishes for
$A^2-H^2=\frac{1}{2}\sqrt{H^2+\frac{1}{16}\hbar^2}+\frac{1}{8}\hbar^2\sim
\frac{1}{2}H\hbar$, which corresponds to symmetric fluctuations, but
increases very rapidly for slightly larger values and can become
arbitrarily large (or very nearly one for the upper sign).  For very
large values one needs a large difference $A^2-H^2$, but even
$A^2-H^2\sim H\hbar$, which is not an extreme value compared to
$\frac{1}{2}H\hbar$ of the symmetric case, gives
$|1-c_3/c_4|=2\sqrt{3}/2-\sqrt{3}\approx 13$. As illustrated in
Fig.~\ref{AsymmFluct}, the asymmetry parameter $|1-c_3/c_4|$ increases
very steeply from zero especially for large values of $H$, i.e.\
massive universes.\footnote{Inhomogeneous situations do not distribute
  the total energy content of matter in a single homogeneous patch and
  thus refer to smaller values of $H$. This could lead to results less
  sensitive to the precise value of fluctuations, but it would also
  put the bounce into a much stronger quantum regime or possibly
  eliminate it altogether; see also \cite{BouncePert}.}

\section{Interpretation}

Uncertainty relations do provide a limit on the asymmetry provided
that the size of fluctuations is known since the covariance and thus
squeezing is limited for given fluctuations. One may thus hope to find
a reliable bound on present fluctuations, by which one could restrict
the squeezing of the state of the universe and thus also the pre-big
bang fluctuations, which would be remarkable.  But how can we find a
strong upper bound on current fluctuations in the state of the
universe?

Density fluctuations are tiny, and were even smaller at the time of
decoupling as we can see them through the cosmic microwave
background. These are classical fluctuations in a very nearly
classical phase, and quantum fluctuations must be even smaller. This
cannot give a good estimate for our purposes, however, because these
are fluctuations of inhomogeneities. They were already assumed to
vanish in the isotropic model used. By putting an observational bound
on something already assumed to vanish one can certainly not derive
anything new. Fluctuations we are dealing with are quantum
fluctuations even of the isotropic variables of a universe.

Moreover, if we wanted to use real observations to restrict the state
of the universe before the big bang we would have to consider
decoherence. Decoherence is certainly acting in the real universe and
is usually understood as making quantum states more classical by
interactions with a large environment of weakly interacting degrees of
freedom. The huge number of degrees of freedom it requires are
also ignored in the model; to compare observations with results in the
model we would thus have to factor out decoherence processes. 

These considerations bring us back to the introductory remarks: We can
use a solvable model as the most optimistic option. Here, one can
explicitly compute the general behavior of states and how some properties
(e.g.\ before the big bang) affect others (e.g.\ those after the big
bang). Making this model more realistic can only complicate the
derivation of any such properties, for which decoherence would be an
example. Even if we take the model for real, there are limitations on
knowledge of the precise form of the pre-big bang state, such as its
fluctuations. The model clearly tells us that, using observations only
after the big bang, we would have to determine the precise squeezing
of the state to extrapolate all the way to the state before the big
bang. Doing this would require so much control that, even disregarding
decoherence, it should be considered hopeless, the existence of upper
bounds on squeezing for given fluctuations notwithstanding. 

What could one infer about the pre-big bang state in a hypothetical
harmonic universe which follows the solutions provided here and does
not show decoherence? In principle this should give one access to
properties of the wave function. Quantum variables such as $G^{vv}$
contain the pre-bounce quantities only with tiny suppression factors
$e^{-2\phi}$ and thus do not provide insights. One would have to
measure current fluctuations very precisely to infer the state
parameters. Assuming that the state is coherent, (\ref{relasymm})
tells us what has to be measured to determine $\delta$. (If the state
cannot be assumed coherent, changes between the pre- and post-bounce
phase would be much more pronounced.)

We thus need to find the three parameters $A$, $H$ and $\Delta H$ from
observations. This is subtle because the asymmetry depends on $A$ and
$H$ only by the ratio $H/A$, which for purposes here can be assumed to
be just unity, and $A^2-H^2\sim O(A\hbar)$. The latter is important
for the precise magnitude of $\delta$, but arises as a small value
from a near cancellation between two large quantities.
Fig.~\ref{AsymmFluct} clearly illustrates how sensitive the asymmetry
is to small changes in the parameters especially for the realistic
case of large $H$.
To see the origin of this quantity it is important to consider a
solvable model with precise access to dynamical coherent states.

The Hamiltonian $H$ quantifies the matter content and can possibly be
measured well. Similar, matter fluctuations $\Delta H$ can be granted
to be under good observational control. Difficulties arise for the
parameter $A$. It determines the size $v$ of the universe at the
bounce, but this can hardly be used for an observational
determination. Alternatively, one could use the present $v$ and
eliminate the factor $e^{\phi}$, the ratio of the present size the
bounce size, through the age of the universe. The internal time $\phi$
is not proper time $\tau$, but they can be related through
$\tau(\phi)= -A^{3/2}H^{-1}\int^{\phi} \cosh^{3/2}(z)\md z$
\cite{BounceCohStates}. Both potentially observable quantities, $v$
and $e^{\phi(\tau)}$, are huge and provide $A$ in their ratio.
Relative uncertainties in each measurement will provide large absolute
uncertainties for $A$, and yet we need to produce a subtle
cancellation with $H$ to find the precise $\delta$.

This prevents a sufficient observational control on the state in order
to determine its pre-bounce fluctuations, even though the evolution is
deterministic. We have presented details only for the moments of order
two, but all orders behave similarly. There are thus infinitely many
variables lacking for a precise knowledge of the pre-bounce state, nut
just the fluctuation.  Thus, the way the classical singularity is
removed in loop quantum cosmology may present a well-defined universe
scenario without divergences of energy density, but it does not allow
us to know precisely what happened before the big bang: the past is
shrouded by cosmic forgetfulness \cite{BeforeBB}.

\section*{Acknowledgements}

The author thanks Abhay Ashtekar, Alex Corichi and Parampreet Singh
for discussions, and an anonymous referee of \cite{BeforeBB} for
pointing out the article \cite{Uncertain}. This work was supported
in part by NSF grant PHY06-53127.


\end{document}